\newcommand {\be}{\begin{equation}}
\newcommand {\ee}{\end{equation}}
\newcommand {\la}{\label}
\newcommand {\bea}{\begin{eqnarray}}
\newcommand {\eea}{\end{eqnarray}}
\newcommand {\bi}{\bibitem}
\documentstyle[aps,prl,epsf]{revtex}
\begin{document}
\draft
\twocolumn[\hsize\textwidth\columnwidth\hsize\csname@twocolumnfalse%
\endcsname
%\rightline{\small published in \prl {\bf 80}, 1694 (1998); cond-mat/9709013}
\title{
Conductance of a Mott Quantum Wire
}
\author{Oleg A. Starykh and Dmitrii L. Maslov}
\address{
Department of Physics, University of Florida,
P. O. Box 118440, Gainesville, FL 32611-8440
}
\maketitle
\begin{abstract}
We consider transport through a one-dimensional
conductor subject to an external periodic
potential and connected to non-interacting
leads (a \lq\lq Mott quantum wire\rq\rq\/).
For the case of a strong periodic potential,
the conductance is shown to jump from
zero, for the chemical potential lying within the
Mott-Hubbard gap, to the non-interacting value
of $2e^2/h$, as soon as the
chemical potential crosses the gap edge.
This behavior is strikingly different
from that of an optical conductivity,
which varies continuously with the
carrier concentration. 
For the case of
a weak potential, the perturbative correction to 
the conductance  due to Umklapp scattering is absent away from half-filling.
\end{abstract}
%---------------
%---------------
]
%\narrowtext
%%%%%%%%%%%%%%%%%%%%%%%%%%%%%%%%%%%%%%%%%%%%%%%%%%%%%%%%%%%%%%%%%%%%%%%
%\section{Introduction}
%%%%%%%%%%%%%%%%%%%%%%%%%%%%%%%%%%%%%%%%%%%%%%%%%%%%%%%%%%%%%%%%%%%%%%%
It has long been understood that
the result of a transport measurement
depends on the measurement procedure \cite{szafer}.
The most famous example is perhaps a ballistic
one-dimensional (1D) wire.
The real part of its  {\it{conductivity}} has a familiar
Drude peak at $\omega=0$, indicating
an obvious fact that an ideal system
exhibits no resistance to the
stationary current flow. On the other
hand, the same wire being attached
to the reservoirs exhibits a finite (two-probe)
{\it conductance} of $2e^2/h$.
The importance of the measurement procedure
has been re-emphasized in recent work
\cite{safi_prb,ms,ponom_prb},
which have shown that the
{\it conductance} of a Luttinger-liquid
wire attached to the reservoirs via non-interacting leads remains
at its non-interacting value of $2e^2/h$,
regardless of the interactions in the wire, 
although the {\it conductivity} of an infinite
wire is indeed renormalized by the interactions.

In this work, we focus our attention
on transport through another system
whose properties are generally believed
to be strongly affected by the
interactions, {\it i.e.}, on
an 1D conductor subject to an
external periodic potential.
In such a system, Umklapp scattering
of electrons leads to an opening
of the Mott-Hubbard gap, so that
the system is an insulator at half-filling.
Away from half-filling some
conduction occurs. Almost all we know
about transport through
such a system \cite{schulz,giamarchi,mori} is valid for a bulk sample (except
for very recent studies \cite{kawakami,raikh,ponom}), whose conductivity
is assumed to be measured
in a {\it contactless} way, {\it e.g.}, via
electromagnetic losses.
The main question
we are asking in this work is: how does
a (doped) Mott insulator conduct when
being attached to non-interacting leads?

For the sake of concreteness, the system
we have in mind will be taken as a quantum wire
subject to a periodic gate potential (cf. Fig.~1a),
similar to that fabricated in the experiments
\cite{LeoK,Tarucha}.
For brevity, we shall refer to such
a system as to a \lq\lq Mott quantum wire\rq\rq\/.
In what
follows, we consider both
the cases of strong and weak periodic
potential 
(compared to the Fermi energy in the wire).
Our main finding is
that {\it as long as a Mott quantum wire is conducting,
it is an ideal conductor}. This
conclusion is qualified as follows. 
i)~In the case of a {\it strong periodic potential}, the wire is
an insulator whenever the chemical
potential lies within the Mott-Hubbard
gap, regardless of the measurement
procedure. However, as the chemical potential
crosses the gap edge,
the conductance
jumps from zero to its non-interacting
value, $2e^2/h$. (In what follows, we will put
$\hbar=1$ everywhere except for the
results for the conductances.)
ii)~ If the system is not at half-filling, the case of a {\it weak
periodic potential}
can be treated
via the perturbation theory, which shows
that the {\it dc} conductance
of the wire is not affected by Umklapp
scattering. 
%%%%%%%%%%%%%%%%%%%%%%%%%%%%%%%%%%%%%%%
%\section{Strong periodic potential.}
%%%%%%%%%%%%%%%%%%%%%%%%%%%%%%%%%%%%%%%

(i)~{\it Strong periodic potential.} In this case a wire
is effectively split into individual
\lq\lq sites\rq\rq\/ (dots) connected
via tunneling of electrons. We assume
that there is about one electron
per site on average, {\it i.e.}, the system
is close to half-filling.
In the presence of Umklapp scattering,
a Mott-Hubbard gap opens up at $k=\pm \pi/2a$, where $a$ is the period of the
potential. 
We begin by considering an infinite
wire without leads.
This situation is best described
by the tight-binding model of a Hubbard type
\bea
H&&=-\frac{t}{2}\sum_{s,n}(\psi^{\dag}_s(n) \psi_s(n+1) + \psi^{\dag}_s(n+1)
\psi_s(n))\nonumber\\
&& - \mu\sum_{s,n}\psi^{\dag}_s(n)\psi_s(n) + g\sum_n 
\psi^{\dag}_{\uparrow} \psi_{\uparrow} \psi^{\dag}_{\downarrow}
\psi_{\downarrow}.
\la{hubbard}
\eea
Linearizing the spectrum around $\cos(k_F a)=-\mu/t$, expanding the fermion
operators in right and left movers, and taking the continuum limit,
we arrive at the following bosonized Hamiltonian for the charge
degrees of freedom
\bea
H&&=\frac{1}{2}\int dx [v_F(\partial_x\theta_{\rho})^2 + (v_F + ga/\pi)
(\partial_x\phi_{\rho})^2\nonumber\\
&& - \frac{g}{a \pi^2}\cos(\sqrt{8\pi}
\phi_{\rho} + q_0 x) + 2gn\sqrt{\frac{2}{\pi}}\partial_x \phi_{\rho}],
\la{bos}
\eea
where $q_0=4k_F - 2\pi/a$, $n=\frac{1}{2}\sum_s<\psi^{\dag}_s(n)\psi_s(n)>$
is an average number of carriers per site, and
$v_F=ta\sin(k_F a)$.
Shifting the $\phi_{\rho}$ field to remove the $q_0x$ term under the cosine
shifts the  chemical potential:
$\mu\to\bar{\mu}=gn - (v_F + ga/\pi)(k_F - \pi/2a)$. 
To treat this Hamiltonian in the strong coupling limit $g \gg v_F$,
we map it onto a system of spinless fermions \cite{emery,schulz}.
Introducing new boson fields 
$\varphi=\sqrt{2}\phi_{\rho}, \theta=\theta_{\rho}/\sqrt{2}$,
which preserve commutation relations, we re-write the  Hamiltonian (\ref{bos})
in terms of new
right- (left-) moving spinless fermions $\psi_{+} (\psi_{-})$ 
\bea 
H=&&\int dx\Big(v:\psi_{+}^{\dag}(-i\partial_x)\psi_{+}
+\psi_{-}^{\dag}(i\partial_x)\psi_{-}:\nonumber\\
&& + \bar{\mu}:\psi_{+}^{\dag}\psi_{+}
+\psi_{-}^{\dag}\psi_{-}: + \frac{ga}{2\pi\alpha}(\psi_{+}^{\dag}\psi_{-}
+\psi_{-}^{\dag}\psi_{+})
\nonumber \\
&&+g_{res}:\psi_{+}^{\dag}\psi_{+}:
:\psi_{-}^{\dag}\psi_{-}:\Big),
\la{ferm}
\eea
where $v=\frac{5}{4}v_F + \frac{ga}{4\pi}, g_{res}=\frac{1}{2}(ga - 3\pi v_F)$,
$\alpha$ is the short-distance cut-off, 
and left-moving fermions were unitarily transformed
$\psi_{-} \rightarrow i\psi_{-}, \psi_{-}^{\dag} \rightarrow 
-i\psi_{-}^{\dag}$. The last term represents the residual interaction
between fermions. The chain of transformations $\phi_{\rho} \rightarrow
\varphi \rightarrow \psi_{\pm}$ describes a change from the charge density
wave description to the charge soliton one. Under this transformation,
the density of charge fluctuations becomes $\rho(x)=\sqrt\frac{2}{\pi}
\partial_x \phi_{\rho}=
\frac{1}{\sqrt{\pi}} \partial_x \varphi=\sum_{s=\pm}:\psi_{s}^{\dag}\psi_{s}:$
and the current $j(x)=-v_F \sqrt\frac{2}{\pi}\partial_x \theta_{\rho}=
2v_F:\psi_{+}^{\dag}\psi_{+} - \psi_{-}^{\dag}\psi_{-}:$.
Following Schulz \cite{schulz}, we diagonalize the quadratic part of
(\ref{ferm}) and find two bands of excitations with dispersion
$\omega_k=\bar{\mu} \pm \sqrt{v^2k^2 + \bar{g}^2}$, where 
$\bar{g}=\frac{ga}{2\pi\alpha}$ is the Mott-Hubbard gap.
If $\bar{\mu}$ crosses the top of the lower band,
then the density of carriers (holes) is $\nu=\sqrt{\bar{\mu}^2 - 
\bar{g}^2}/(\pi v)$ for $\bar{\mu} > \bar{g}$, and zero otherwise.
Close to half-filling, we again
linearize the spectrum around the hole Fermi points $k_c =\pm\nu/\pi$
and represent the low-energy charge excitations
in terms of $new$ right- (left-) moving fields $\beta_1 (\beta_2)$ with
momenta close to $k_c$. In doing that, one finds that the residual
interaction term renormalizes to $g_{res}(\frac{vk_c}{\bar{\mu}})^2
\int dx :\beta^{\dag}_1 \beta_1::\beta^{\dag}_2 \beta_2:$, whereas
the part 
\be
H_0=\int dx \frac{v^2k_c}{\bar{\mu}}\Big(\beta^{\dag}_1 (-i\partial_x)
\beta_1 + \beta^{\dag}_2 (i\partial_x)\beta_2 \Big)
\la{ferm2}
\ee 
describes free propagation with the renormalized velocity
$\bar{v}=\frac{v^2k_c}{\bar{\mu}}=v\sqrt{1 - \bar{g}^2/\bar{\mu}^2}
\theta(\bar{\mu} - \bar{g})$. As half-filling is
approached, {\it i.e.}, as $k_c \rightarrow 
0$, the residual interaction vanishes faster than the renormalized velocity
due to the additional power of $k_c$, and thus can be neglected
altogether. The bosonization of free Hamiltonian (\ref{ferm2}) is 
straightforward, and we find for the low-energy action
\be
S_0=\frac{1}{2}\int dt\int dx [\frac{1}{\bar{v}}(\partial_t\bar{\varphi})^2 
- \bar{v}(\partial_x\bar{\varphi})^2],
\la{act2}
\ee
while the charge density becomes $\rho(x)=\partial_x
\bar{\varphi}/\sqrt{\pi}$ and the current $j(x)=-2v_F
\sqrt{1 - \bar{g}^2/\bar{\mu}^2} \partial_x \bar{\theta}/\sqrt{\pi}$.
The continuity equation $\partial_t \rho(x) + \partial_x j(x)=0$
then leads to $v=2v_F$.
The non-local conductivity $\sigma(\omega,q)$
can now be calculated with the help of either fermionic 
[Eq.(\ref{ferm})] or bosonic [Eq.(\ref{act2})] formalisms. 
Both approaches give for the optical conductivity
$\sigma(\omega,0)=4\pi e^2 v_F^2 \nu\delta(\omega)/{\bar \mu}$
\cite{giamarchi,mori}. Notice that the Drude weight 
goes to zero linearly with the density of carriers.
On the other hand, the conductance $G$ (defined
as the ratio of the current to the voltage which is
applied to a finite segment of otherwise uniform wire)
is related to the static conductivity
$\sigma(0,q)$,
which in
our case takes the form
$\sigma(0,q)=e^2\delta(q)\theta(\bar{\mu} - \bar{g})/2\pi$.
Correspondingly, $G=(e^2/h)\theta(\bar{\mu} - \bar{g})$, which
for $\bar{\mu}>\bar{g}$
coincides with the conductance  of noninteracting spinless
electrons
(in agreement with $K_{\rho} \rightarrow 1/2$ for 
$\nu \rightarrow 0$ \cite{schulz})
and drops abruptly to zero in the insulating phase, when $K_{\rho}=0$.
To be more precise, we have to recall that the
continuum approximation works only when the average
distance between the carriers $1/\nu$ is smaller than
the wire length $L$; thus a narrow region, in which
${\bar \mu}$ is so close to ${\bar g}$
that $\nu L\leq 1$, has to be excluded.  

To describe a finite Mott wire connected to a non-interacting leads
we use the approach due to  Safi and Schulz \cite{safi_prb,safi}.
The action of charge modes in the leads is that of a free boson field
$\varphi_{\rho}$ with velocity $v_F$, and thus can be obtained from
action (\ref{act2}) by a simple substitution $\bar{v} \rightarrow v_F,
\bar{\varphi} \rightarrow \varphi_{\rho}$.
Observe now that we can describe 
the charge dynamics in the whole system  by a single equation
\be
\frac{1}{v(x)}\partial_t^2\chi - \partial_x(v(x)\partial_x\chi) =0,
\la{eq2}
\ee
where $v(x)=v_F, \chi=\varphi_{\rho}$ in the leads,
and $v(x)=\bar{v}, \chi=\bar{\varphi}$ in the wire.
With this identification we have 
$\rho(x)=\frac{z}{\sqrt{\pi}}\partial_x \chi$ and 
$j(x)=-\frac{z}{\sqrt{\pi}}\partial_t \chi$
throughout the system,
where $z=\sqrt{2}$ in the leads and $z=1$ in the wire.
Let us consider now a transmission of a boson wave from the lead 
to the wire \cite{safi}, corresponding to a transmission
of a single charge:
\bea
&&\chi_{lead}=e^{i(qx - \omega_q t)} - R e^{-i(qx + \omega_q t)},
\nonumber\\
&&\chi_{wire}=T e^{i(q^{\prime} x - \omega_{q^{\prime}} t)}.
\eea
The density reflection coefficient $R=\partial_x \chi_{lead}^{l}/
\partial_x \chi_{lead}^{r}$, where $r$ ($l$) refers to
right (left) moving excitations.
Current continuity $j(-0)=j(+0)$ gives $\sqrt{2}(1 - R)=T$,
whereas the condition $v_F \partial_x \chi(-0)=\bar{v}\partial_x \chi(+0)$
gives $T=1+R$, where we have also used the energy conservation 
$\omega_q=\omega_{q^{\prime}}$, and thus 
$R=\frac{\sqrt{2}-1}{\sqrt{2}+1}$.
Analogous consideration for the wave traveling from the wire to the
lead shows that its reflection coefficient is equal to $-R$.
The system lead-wire-lead can now be modeled as two semi-transparent
mirrors, the sign of the reflection coefficient
depending on whether the wave is coming
from the wire or the lead \cite{safi_prb,safi}.
Summing over  multiple reflections, we obtain the total
charge transmitted through the system  $Q_{trans}=(1 -R^2)
\sum_{n=0}^{\infty} R^{2n}=1$, whereas the reflected charge is
$Q_{refl}=R - (1 -R^2)\sum_{n=0}^{\infty} R^{2n+1}=0$.
Thus we have {\it perfect transmission} and $G=\frac{2e^2}{h}$.
This result is valid as long as there are carriers
in the wire, i.e., away from half-filling.
At half-filling, when there are no propagating excitations
in the wire, the reflection coefficient
is equal to 1 and the conductance is equal to zero
(we neglect here tunneling which is exponentially
small for long enough wires).
Summarizing,
\be
G=\frac{2e^2}{h}\theta(\bar{\mu}-\bar{g})
\la{sc_cond},
\ee
which is twice as large as in the case
of homogeneous interaction strength.
(If the chemical potential lies in the upper band, the argument of the $\theta-$
function in Eq.(\ref{sc_cond}) becomes $\bar{g}-\bar{\mu}$.)
\begin{figure}[htb]
\setlength{\unitlength}{1.0in}
\begin{picture}(2.0,3.0)(0.1,0) 
\put(0.15,0.1){\epsfxsize=3.0in\epsfysize=3.0in\epsfbox{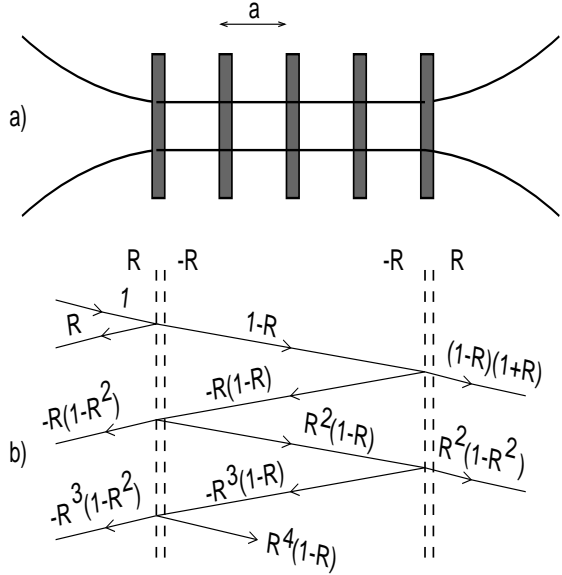}}
\end{picture}
  \caption{
{\bf (a)} Schematic view of a Mott quantum wire;
{\bf (b)} Transmission of the charge boson wave, coming from the left with 
the amplitude 1,
through a Mott quantum wire subject to a strong periodic potential.
Amplitudes of the transmitted and reflected waves for few first consecutive
scatterings are shown.}
\end{figure}
%%%%%%%%%%%%%%%%%%%%%%%%%%%%%%%%%%%%%%%%%%%%%%%%%%%%%%%%%%%%%%%%%%%%%%%%%%%%%%%
%\section{Weak periodic potential.}
%%%%%%%%%%%%%%%%%%%%%%%%%%%%%%%%%%%%%%%%%%%%%%%%%%%%%%%%%%%%%%%%%%%%%%%%%%%%%%

(ii)~{\it Weak periodic potential.}
To see whether the non-perturbative results of
section ({\it{i}}) could be approached in
a perturbative way,
we consider a continuum 1D interacting electron system 
subject to a periodic potential $W(x+a)=W(x)$, as described 
by the Hamiltonian 
\bea
H&&=\int dx \psi_s^{\dag}(x) \left(-\frac{\partial^2_x}{2m}\right) \psi_s(x)
+W(x)\psi_s^{\dag}(x)\psi_s(x)\nonumber\\
&&+\sum_{ss'}\int dx d{x^{\prime}} U(x - x^{\prime}) \psi_s^{\dag}(x)\psi_s(x)
\psi_{s^{\prime}}^{\dag}(x^{\prime})\psi_{s^{\prime}}(x^{\prime}).
\la{H}
\eea
Note that
we are no longer constrained by the half-filling
condition, which was essential in section ({\it{i}}).
In the presence of a periodic
potential, the correct 
single-electron basis is formed
by the Bloch states, which
has to be used in the decomposition
of the interaction term in Eq.~(\ref{H})
into the interaction of left- and right-movers.
We assume that $\mu$
is sufficiently far away from
the Bragg gaps which open 
at $k=\pi l/a, l=\pm 1\dots$.
Under this condition, corrections to the plane-wave basis due to the
periodic potential can be found via the non-degenerate perturbation theory.
In the leading order, 
the effective
coupling constant for
Umklapp scattering takes the form 
$g_U=8m W_{G_0} U(2k_F)/\pi^2G_{0}^2$,
where $W_G$ is the Fourier component
of $W(x)$ and $G_0$ is a reciprocal wavevector
chosen in such a way that
$|4k_F-G_0|$ is minimal.
It is worth noting that $g_U =0$ 
in either a spacially uniform or non-interacting system.

Bosonization of (\ref{H}) leads to the following action
\bea
S&&=\int dx d\tau \Big\{\frac{1}{2K(x)v(x)}
(\partial_{\tau}\varphi_{\rho})^2 + 
\frac{v(x)}{2K(x)}(\partial_x \varphi_{\rho})^2\nonumber\\ 
\!\!&&+\! \frac{g_U(x)}{\alpha^2} \cos(\sqrt{8\pi}\varphi_{\rho}\!+\!q_0 x)
\!\Big\},
\la{action}
\eea
where $q_0=G_0 - 4k_F$
and $K$ characterizes the strength of interactions.
Eq.(\ref{action}) is generalized to the
inhomogeneous case \cite{safi_prb,ms,ponom_prb}, which 
implies that the external periodic potential is applied to the quantum
wire ($|x| \leq L/2$), which is described by the set of
constants $\{v_2, K_2, g\}$. Noninteracting leads with 
$\{v_1=v_F, K_1=1, g=0\}$ 
are attached adiabatically to the both ends of the wire.

 The conductivity is related to the retarded Green's function of the 
field $\varphi_{\rho}$ via the Kubo formula \cite{shankar}
\be
\sigma_{\omega}(x,y)=\frac{-i e^2\omega}{\pi}  
G_R(x,y,\omega).
\label{cond}
\ee
To the second order in $g_U$, one has
\bea
&&\delta G_R(x,y,\omega)=4\pi\int dx_1 dx_2 ~\frac{g_U^2(x)}{\alpha^4}
\cos(q_0(x_1 - x_2))\nonumber\\
&&\times G_R(x_1,x,\omega)
\Big(
G_R(x_2,y,\omega)
F_R(x_1,x_2,\omega)\nonumber\\
&&-G_R(x_1,y,\omega)
F_R(x_1,x_2,0)\Big). 
\la{dG}
\eea
Here the retarded function\\
$F_R(x_1,x_2,\omega)=2\int_0^{\infty} dt
e^{i\omega t} {\text {Im}} F_M(x_1,x_2,it - 0)$ is related to
the imaginary-time correlator of $\cos(\sqrt{8\pi}\varphi_\rho)$ \cite{shankar}
\bea
&&F_M(x_1,x_2,\tau)=\exp\Big\{-8\pi \int_0^{\infty} \frac{d\bar{\omega}}{2\pi}
\Big(G_M(x_1,x_1,\bar{\omega}) 
\nonumber\\&&
+ G_M(x_2,x_2,\bar{\omega}) - 
2G_M(x_1,x_2,\bar{\omega})\cos{\bar{\omega}\tau}\Big)\Big\}.
\eea
Let us start with the case of a uniform wire [$K(x),v(x)=const$ for all $x$]
subject to a weak periodic potential applied
to a finite segment of length $L$, {\it i.e.},
$g_U(x)=0$ for $|x|>L/2$. 
We concentrate on the limiting cases of \lq\lq high\rq\rq\/
and \lq\lq low\rq\rq\/ frequencies, {\it i.e.},
when $\omega_L\equiv L\omega/v\gg 1 (\ll 1)$, respectively, and
assume that $\omega \ll vq_0$.
The difference between these two cases comes about from the 
integration over the
center-of-mass coordinate of the pair $(x_1,x_2)$.
Carrying out this integration,
we obtain
\be 
\delta G_R(x,y,\omega)=\frac{2\pi g_U^2 K^2}{\alpha^4\omega^{2}}\int
_{0}^{\infty}
 ds \cos(q_0s){\cal F}(s),
\la{s_int}\ee
where $s=(x_1-x_2)/2$ and
\bea
&&{\cal F}(s)=
\left\{
\begin{array}{l}
P(0)F_R(s,0) - P(s)F_R(s,\omega); \omega_L\gg 1;\\
\theta(L-s)(L-s)[F_R(s,0)-F_R(s,\omega)]; \omega_L\ll 1.
\end{array}\right.
\nonumber
\eea
Here $P(s)=\sum_{j=\pm}(\frac{i v}{\omega} + |x-y + js|)e^{i\omega|x-y+js|/v}$
and $F_R(s,\omega)=\sin{2\pi K} \int_{|s|/v}^{\infty} dt
 e^{i\omega t}\left(\frac{t_f^2}{t^2 - s^2 /v^2}\right)^{2K}$, 
where $t_f^{-1}\sim E_F$ is the high-frequency cut-off.
Finally, relating the conductivity to the current via 
$I(x,\omega)=\int_{-L/2}^{L/2}dy \sigma_{\omega}(x,y) E_{\omega}(y)$ and
defining the conductance as $G(\omega)=I(L/2,\omega)/V(\omega)$,
we find for the correction to the conductance $\delta G(\omega)=-
(e^2g_U^2/hv^2)\times\gamma(\omega)$, where
\bea
\gamma(\omega)=
\left\{
\begin{array}{l}
C_>\cos(\omega_L) (q_0\alpha)^{4K - 4};
\omega_L\gg 1;\la{dgu}\\
C_< \frac{1 -\cos(q_0 L)}{(q_0\alpha)^2} 
(\omega t_f)^{4K -2};
\omega_L\ll 1,
\end{array}
\right.
\eea
where $C_>=2^{1-4K}(2K - 1) \sin(2\pi K) \Gamma^2(1 - 2K) K^2$
and $C_<= \frac{2\pi}{\Gamma(4K)}$.
The high-frequency result could
have been obtained by assuming that
the periodic potential is applied
over infinite length (as in, {\it e.g.}, Ref.~\cite{giamarchi}),
{\it{i.e.}} it is a \lq\lq bulk\rq\rq result.
The low-frequency limit shows a remarkable
feature: away for half-filling, the correction to the conductance
of a finite-length wire is absent. Indeed, 
for $q_0 L \gg 1$, where our results are only valid,
 $q_0^{-2}(1 -\cos(q_0 L)) \to\pi L^2 \delta(q_0 L)/2$,
{\it i.e.}, the periodic potential does not affect the conductance 
for $q_0\neq 0$.

We now return to the original problem of interest: a quantum wire of length $L$
with parameters $\{K_2,v_2,g_U\}$
connected to the leads with parameters $\{K_1,v_1,0\}$.
In the high-frequency limit, the correlation
function $F_R$ reduces to its short-time and  short-distance
asymptotic form, which coincides with that
of a uniform wire with parameters $K_2,v_2$.
Therefore, the result is given by the top line
in Eq.~(\ref{dgu}) with
$K\to K_2,v\to v_2$.
In the low-frequency limit, the $t-$ integration is determined by the
asymptotics of $F_R$ at $t \sim 1/\omega \gg t_L=L/v_2$.
By using
an explicit form of $G_M$ for an inhomogeneous
Luttinger liquid \cite{ms,ponom_prb}, one can see that for such long times
the time dependence of $F_R$ is determined by
the leads: 
$F_R(t)\sim (t_f/t_L)^{4K_2}{\text {Im}}\left\{t_L/(it-0)\right\}^{4K_1}$.
Concentrating on the case of noninteracting leads ($K_1 =1$), 
we find 
\be
\delta G(\omega)=-\frac{\pi^2}{3}\frac{e^2}{h}\frac{g_U^2}{v^2_2} 
\Big(\frac{\alpha}
{2L}\Big)^{4K_2 - 4} \delta(q_0 L) (\omega t_L)^2.
\la{dG2}
\ee
Thus away from half-filling, there is no perturbative correction to the
conductance due to a weak
periodical modulation of the quantum wire connected to the leads
(cf. also Ref.~\cite{ponom}).
This result is
a perturbative analog of
the result (\ref{sc_cond}) from the previous section.

Absence of the conductance correction
in the perturbation theory can be understood as follows.
For $\omega \ll v/L$, boson field
$\varphi_{\rho}$ varies slowly  on the scale of $L$ and hence
the cosine-term in (\ref{action}) can be approximated as 
$g_U \cos(\sqrt{8\pi}\varphi_{\rho}(0,\tau))\int_{-L/2}^{L/2} dx \cos(q_0x) \sim
\pi gL\delta(q_0L)\cos(\sqrt{8\pi}\varphi_{\rho}(0,\tau))$, from which
it is clear that this perturbation is effective only at half-filling.
It can be shown that both $1/K(x)$ and
$1/v(x)$ acquire singular corrections proportional to $\delta(x)$, {\it i.e.},
their local values at $x=0$ tend to infinity. 
The scaling dimension of the $\cos(\sqrt{8\pi}\varphi_{\rho}(0,\tau))$ operator
with $K$ and $v$ vanishing  at
$x=0$ is equal to zero. 
Thus this operator
is relevant and requires a non-perturbative
treatment which was carried out in the previous section.

We thank Vadim Ponomarenko for elucidating
some points of Ref.~\cite{ponom} for us
and  Seigo Tarucha for several
inspiring discussions. This work
was supported by the NSF (DMR-9703388).

In the process of writing of this paper,
we learned of recent papers by Odintsov, Tokura, and Tarucha
(\prb {\bf 56} R12729 (1997)) and by Mori, Ogata, and Fukuyama
(J. Phys. Soc. Jap. {\bf 66}, 3363 (1997)).
Although the $\omega$-dependence of our result (\ref{dG2}) 
transforms into the temperature-dependence by Odintsov et al.
upon $\omega\to T$,
our $L$-dependence contains an extra 
factor $(\alpha/L)^{4K_2-4}$, which manifests
the amplification of Umklapp scattering due to
the multiple reflections
of electrons at the wire-lead interfaces.
Our result for the case of a strong potential Eq.~(\ref{sc_cond})
coincides with that by Mori et al.;
we would like to thank M.\ Mori for his clarifying comments.

\end{document}